# 9-1-1 DDoS: Threat, Analysis and Mitigation

Mordechai Guri, Yisroel Mirsky, Yuval Elovici
{gurim,yisroel, elovici}@post.bgu.ac.il
Ben-Gurion Univeristy of the Negev
Cyber-Security Research Center

*Abstract* - **The 911 emergency service belongs to one of the 16 critical infrastructure sectors in the United States. Distributed denial of service (DDoS) attacks launched from a mobile phone botnet pose a significant threat to the availability of this vital service.**

**In this paper we show how attackers can exploit the cellular network protocols in order to launch an anonymized DDoS attack on 911. The current FCC regulations require that all emergency calls be immediately routed regardless of the caller's identifiers (e.g., IMSI and IMEI). A rootkit placed within the baseband firmware of a mobile phone can mask and randomize all cellular identifiers, causing the device to have no genuine identification within the cellular network. Such anonymized phones can issue repeated emergency calls that cannot be blocked by the network or the emergency call centers, technically or legally. We explore the 911 infrastructure and discuss why it is susceptible to this kind of attack. We then implement different forms of the attack and test our implementation on a small cellular network. Finally, we simulate and analyze anonymous attacks on a model of current 911 infrastructure in order to measure the severity of their impact. We found that with less than 6K bots (or $100K hardware), attackers can block emergency services in an entire state (e.g., North Carolina) for days. We believe that this paper will assist the respective organizations, lawmakers, and security professionals in understanding the scope of this issue in order to prevent possible 911-DDoS attacks in the future.**

## 1. INTRODUCTION

The '911' emergency number was instituted in the US in 1968 in response to the need for a universal and effective method of reporting emergencies. Over the years the system has evolved, and in 1999 the US government enacted the Wireless Communications and Public Safety Act. This federal law mandated the use of 911 as a universal emergency number and "enhanced 911" (E911) as the base *technology* for handling calls from wireline and wireless phones. The E911 network provides dedicated infrastructure for routing and connecting 911 calls to the nearest public call center. These call centers are referred to as public safety answering points (PSAP).

Today, 911 services are a part of the United States' 16 critical infrastructure sectors [1]. Its availability is vital to the population of the United States.

### 1.1 DDoS Attacks on 911

A DDoS attack launched from a mobile phone botnet is a significant threat to the availability of 911 services. In this attack, frequent fraudulent calls are made to 911 by a botnet comprised of many mobile phones. Many PSAPs work at full capacity [2] and cannot handle this large volume of calls. Moreover, this call volume can disrupt the telephony network itself, preventing legitimate 911 calls from ever reaching a PSAP. This was evident during the 9/11 terror attack which, in effect, caused the population to generate a DDoS attack on New York City's telephony network by collectively dialing 911 [3].

In 2015 over 90% of American adults owned a cell phone, and 64% of the devices were smartphones [4]. An attacker that recruits even a fraction of these devices to a botnet would give this attacker has the potential to deny 911 services to an entire state, or possibly the entire country.

This attack currently affects both the service and the client in the following ways: 1) Generally, PSAPs have no built-in way of blacklisting callers. Therefore, in the face of a large attack, they would have no choice but to answer each and every call. 2) Even with a blacklisting system in place, the owner of an infected device would be blocked from legitimately receiving emergency services, even in a time of need.

### 1.2 An Anonymized DDoS Attack on 911

In 1996 the Federal Communications Commission (FCC) issued the E911 First Report and Order which required wireless providers to forward 911 calls to a PSAP regardless of caller validation:

*"The basic 911 rules require wireless carriers to transmit all 911 calls to a Public Safety Answering Point (PSAP) without regard to validation procedures intended to identify and intercept calls from non-subscribers. Under the rules, therefore, both subscribers and non-subscribers can dial 911 and reach emergency assistance providers without having to prove their subscription status."*

~FCC WIRELESS 911 REQUIREMENTS [5]

A DDoS attack launched from a mobile phone botnet can exploit this ruling in order to make its attack more difficult to mitigate. If a bot randomizes the device's cellular identifiers, it becomes impossible to blacklist its 911 calls.

In this paper we expose and analyze this new threat by proving its feasibility and by measuring its potential impact via simulations. We found that only 6,000 infected devices are enough to severely harm the availability of the 911 services of a US state. We also found that some device-level and network-level countermeasures can help in mitigating this threat.

### 1.3 Contributions

The following are the contributions of this paper:

- We identify a new threat to the availability of emergency services: an anonymous, unblockable 911-DDoS attack from mobile phones. We present how this attack is perpetrated and discuss the bots involved.
- We implement bots that can carry out this attack, and test them on a small cellular network in order to demonstrate the attack's feasibility. By describing the behavior and implementation of these samples, we provide the tools and knowledge to prevent potential attacks in the future.

- We gauge the severity of the attack and analyze the weaknesses of the E911 networks by simulating the attack on a reconstruction of actual E911 infrastructure. The simulations are based on real call volume statistics, network topologies, and configurations which we have collected from various published surveys, reports, statistics, and documentation. We analyze possible weaknesses of the current E911 network and measure the number of bots required to accomplish an attack.

We believe that this paper will assist the respective organizations, lawmakers, and security professionals in understanding the scope of this issue. It is our intent that the analysis, simulations, and models will provide these stakeholders with a means of detecting and preventing possible attacks in the future.

The remainder of this paper is structured as follows: In section 2 we review related work. In section 3 we provide a technical overview of the cellular network and E911 infrastructure as relevant to this paper. In section 4 we introduce the adversarial attack model and the bot types. In section 5 we present the design and implementation. In section 6 we provide an in depth analysis of the attacks at the state and country level. In section 7 we present countermeasures, and we conclude in section 8.

## 2. RELATED WORK

**Denial of Service (DoS).** A DoS attack is an attack that targets the availability of a service. DDoS attacks are DoS attacks which are usually launched from a group of devices (botnet), each of which is infected with a malicious agent (bot). A wide range of studies have been published on this subject, demonstrating different types of DDoS attacks [6] and related detection [7] and defense [8] techniques. A Telephony Denial of Service (TDoS) attack is a special type of DoS attack that targets the telephone line's availability. However, TDoS attacks have been mainly studied within the context of VoIP telephone systems [9]. Throughout this paper we will refer to Telephony DDoS with the more common term DDoS. DDoS attacks are considered one of the major threats and most challenging problems of today's cyber-security world [10]

**Mobile Phone Botnets**. With the rise of smartphone popularity, botnet authors have been increasing their efforts on mobile devices [11]. A detailed survey on the timeline of mobile botnets and their characteristics is provided by Nigam [12]. In 2010, Mulliner and Seifert presented a cellular botnet architecture and evaluated it with several practical implementations [13]. In 2011 Taining el al presented "Andbot," a stealthy, resilient, and low-cost botnet designed for smartphones [14]. Mobile botnets have also been used to dial premium numbers send premium SMSs, or conduct spam campaigns. A thematic taxonomy on smartphone botnet attacks is given in [15].

In September 2015, 650,000 Chinese smartphones launched a DDoS attack against websites by collectively issuing over 275,000 HTTP requests per second [16]. This event demonstrated the threat of DDoS attacks launched from mobile phone botnets. Traynor et al demonstrated how a mobile botnet composed of ~11K compromised mobile phones could be used in a DDoS attack targeting the core of a cellular network, an attack which could degrade cellular service by 93% on a regional scale [17].

**911-DDoS.** DoS attacks on emergency call centers have been publicly mentioned in [18] [19]. In 2013, the US Department of Homeland Security (DHS) and Federal Bureau of Investigation (FBI) issued an alert stating that various public services may be vulnerable to DDoS attacks [20]. This warning was triggered due to a DoS attack launched against the administrative line of a PSAP (not the 911 emergency line). In 2014, Dameff et al presented "Hacking 911" at DEF CON and provided a general description of attack vectors on 911 services [21]. The authors broadly discussed line-cutting, cell phone jamming, and DDoS attacks. [21] does not review the 911-DDoS attack in depth and lacks discussion on the implementation, analysis, and mitigation of this threat.

Table 1 compares existing methods of DDoS attacks on 911. As can be seen, attacks launched from landlines, mobile phones, and pre-paid mobile phones expose the phone number, SIM identifier (IMSI), and device-identifier (IMEI) to the cellular network. The bot types presented in this paper can mask these identifiers during the attack, hence making them resilient to blacklisting and blocking.

To the best of our knowledge, there has been no academic research on the threat of anonymous 911-DDoS attacks launched from a mobile phone botnet.

## 3. TECHNICAL BACKGROUND

In this section we provide background information on the cellular and E911 networks, as relevant to this paper.

### 3.1 Cellular Network

2G, 3G, and 4G are three generations of mobile networks (referred to as GSM, UMTS, and LTE respectively), based on different standards, network architecture, and protocols. Throughout this paper, we will refer to the elements and specifications common to all three generations in common terms.

**IMSI** (International Mobile Subscriber Identity). The IMSI is a unique identification number across all cellular networks which is used to identify a mobile subscriber. This number is stored in the SIM card as a 15 digit number. For privacy reasons a randomly generated TMSI (Temporary Mobile Subscriber Identity) is often used instead of the actual IMSI

**Table 1. Comparisons of DDoS attacks on 911.**

| | | *Identifiers* | | |
|---|---|---|---|---|
| | | **Phone Number** | **SIM Identifier** | **Device Identifier** |
| *Attack Origin* | Landlines **[9]** | Yes | N/A | N/A |
| | Mobile phones **[21]** | Yes | Yes | Yes |
| | Pre-paid **[66]** | Yes | Yes | Yes |
| | Our work | No | No | No |

for authentication, location updates, paging, call requests, and other activities.

**IMEI** (International Mobile Station Equipment Identity). The IMEI is a 15 digit number for identifying mobile equipment (e.g., mobile phones and cellular modems) in cellular networks [22]. The IMEI is either stored in the device's firmware or burned into its ROM. The cellular network can request the IMEI of a device during the authentication phase, in order to check it against the Central Equipment Identity Register (CEIR) which blacklists stolen devices.

**NSI** (Non-Service Initialized). A mobile phone registers and unregisters with the network using the *IMSI attach* and *IMSI detach* procedures [23]. When a mobile phone is unable to register to the network, the network considers the device NSI and does not provide the device with data or call services. A device may be NSI when 1) no SIM card is inserted (no-SIM), 2) the subscription associated with the SIM is inactive, 3) the subscriber has not paid his/her bills, 4) the IMSI or roaming subscriber is unknown, or 5) the IMEI has been blacklisted (e.g., a stolen device).

**Emergency Calls.** When a user dials an emergency number, a CM_SERVICE_REQUEST with a parameter indicating the establishment of an emergency call, is sent to the network [23]. In LTE, a special *emergency attach* procedure is required before issuing emergency calls. Emergency calls can be made from any device-network registration state, even the NSI. This is because FCC regulations required all cellular operators to route all emergency calls to their destination, regardless of the mobile phone's available identifiers [5]. Currently, a device must provide at least provide an IMEI in order to receive emergency services [24].

Table 2 summarizes the different states of device registration and the identifiers exposed to the network in each case. As can be seen, even in the NSI modes, the IMSI or IMEI (or both) are sent to the network.

### 3.2 E911 Network

According to the National Emergency Number Association (NENA), at least 99% of the population of the US have access to E911 [25]. Therefore, for the duration of this paper we only consider the E911 infrastructure of the US. In general, when someone dials 9-1-1 the telecom provider connects the call to the E911 network. The E911 network is responsible for routing 911 calls to the PSAP nearest the caller, as well as providing useful information about the caller (e.g., name and location) to the PSAP's receiving call taker. Fig. 1 presents an illustration of an E911 network and its elements. The edges in the network are groupings of trunks (single voice lines).

**Public-Safety Answering Point (PSAP).** PSAPs are call centers which receive 911 calls originating from within their service area. There are two kinds of PSAPs: primary and secondary. Primary *PSAPs are the PSAPs that first receive the 911 call. In some cases,* a primary PSAP may transfer a 911 call to a secondary PSAP when deemed necessary. An example of a secondary PSAP is a poison control hotline.

**Selective Router (SR).** There are special dedicated telephony switches called selective routers which are located between the public telephone networks and the PSAPs. The SRs are responsible for connecting 911 calls from the telephone network to the primary PSAP closest (geographically) to the caller. SRs are owned by the local emergency service organization but are located and managed by a local telecom company.

**Telephone Switch (TS).** When a call is made from a landline phone, it travels over the public switched telephone network (PSTN) to a local switching station sometimes referred to as the central office (CO). COs are either connected to a SR, or in rare cases, are connected directly to a PSAP. When a call is made by a mobile phone, its signal is received by a nearby cell tower whose physical link is managed by a base station controller (BSC). The BSC connects the call to a mobile switching center (MSC or S-GW in 4G) for call management and call switching. Like a CO, a MSC is connected to a SR.

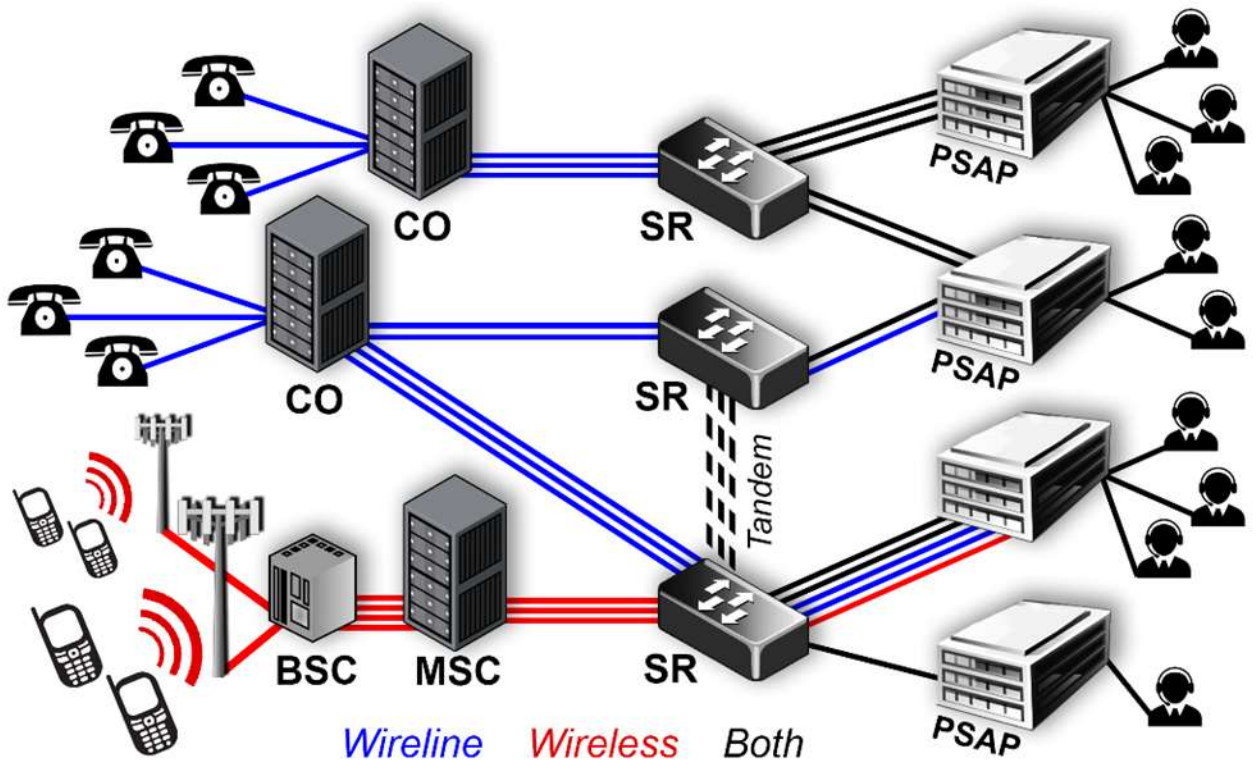

**Figure 1. An illustration of an E911 network topology.**

**Caller Identification and Location Information.** Caller identifiers are provided to the PSAP by the PSTNs and cellular networks. These identifiers are used to retrieve relevant information about the caller from a shared database, called an automatic location information (ALI), between the E911 network and the other networks. For wireless callers, if the device is not registered on the cellular network (i.e., NSI), the call-back number displays zeros or the mobile's IMEI.

**Call Routing.** Call routing in the E911 network works as follows. First, when a TS receives a 911 call, the call is connected with its local SR. If all of the trunks between a TS and a SR are in use, a traffic overflow occurs. In this case, the TS may have an additional SR configured to receive the extra traffic; otherwise the call is dropped. When a SR receives a 911 call from a TS, it attempts to connect the call to the primary PSAP which is closest to the caller's location. To determine the PSAP, the SR has a routing table called a Selective Router Database (SRDB) which links each caller's identifier with a PSAP [26]. The SR then connects the call to the target PSAP. If a traffic overflow occurs between the SR and PSAP, then either: 1) the caller is automatically transferred to an alternate predesignated PSAP (overflow routing), or 2) the caller receives a busy signal and is disconnected. If there is an available trunk but no available call taker, the call is connected, but the caller may hear a call progress-tone until the call is answered.

**Table 2. Different states of device registration in the cellular network, along with the exposed identifiers.**

| *State* | *NSI* | *During registration* | *During emergency calls* |
|---|---|---|---|
| *Registered* | | IMSI,IMEI | IMSI and IMEI |
| *No-SIM* | X | - | IMEI |
| *Non-payed / expired* | X | IMSI | IMSI |
| *Unknown IMSI* | X | IMSI (unknown) | IMSI and IMEI |
| *Blocked IMEI* | X | IMSI, IMEI | IMEI |

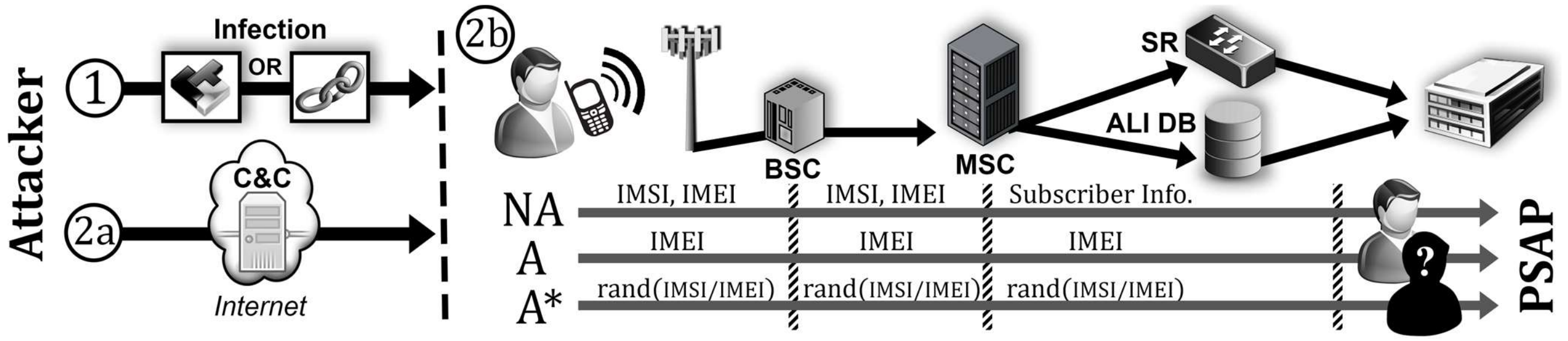

**Figure 2. The adversarial attack model and an illustration of the bot's identifiers carried over the cellular and E911 networks.**

Both wireline and wireless calls are routed over the same SR-PSAP trunks with no distinction. However, some SR-PSAP links have dedicated trunks for wireline or wireless calls [27]. The reason for doing this is to ensure that, for example, a rush of wireless calls do not totally block out wireline calls. During a call, a PSAP may transfer the call to a different PSAP. If the other PSAP is connected to a different SR, a tandem connection may be used if available.

## 4. ADVERSARIAL ATTACK MODEL

The adversarial attack model (Fig. 2) is as follows: The attacker builds a botnet of mobile phones by infecting devices with malware (Fig. 2, step 1). At the appropriate time, the attacker signals the botnet to continuously call 911 (Fig. 2, step 2a). This command is signaled via the attacker's command and control server (C&C) located somewhere on the Internet or by sending covert SMSs to the bots. The attacker can infect the mobiles by using common mobile infection vectors: app markets, email attachments, compromised websites and malvertising campaigns, and malicious SMS/MMS [28]. Supply chain attacks, where devices were found to be pre-installed with malicious code, have been found in the wild [29].

There two kinds of anonymous bots: anonymous ($A$), and persistent anonymous ($A^*$). We refer to the type of bot that is not anonymous as ($NA$). These bots differ in the identifiers they provide to the cellular network when making a 911 call (Fig. 2, step 2b). Anonymous bots cannot easily be identified and blocked by the network. These bots strengthen the impact of the DDoS attack, since they exist longer in the botnet. We will now describe each of these types of bots and their implications.

**Non-Anonymized ($NA$).** $NA$ is a type of bot which makes 911 calls from the device's main OS (e.g., Android and iOS). When doing so, the identifiers (IMSI/IMEI) are exposed to the cellular network as in any non-emergency call. The subscriber information reaches the PSAP where it can then be used to assist in blocking the DDoS device.

**Anonymized ($A$).** $A$ is a type of bot which hides its subscriber information (IMSI) from the network in order to make 911 calls, and is implemented within the firmware of the device's baseband processor. These bots accomplish this by virtually entering No-SIM state, a state where only the IMEI is exposed to the network [24]. In some cases an IMEI can be traced back to the owner, but it is a long and difficult process. In addition, locating and collecting $A$ type bots from the field is more difficult than for $NA$ type bots because the location-detection of NSI devices can be inaccurate [30].

**Persistent Anonymized ($A^*$).** $A^*$ is a type bot that randomizes the identifier (IMSI or IMEI) provided to the cellular network, and is implemented within the firmware of the device's baseband processor. $A^*$ cannot be blocked by use of a blacklist implemented in the cellular network, E911 network, or at the PSAP. This is because the bot changes the identifier before every call. We provide two variations of **($A^*$)** implementation, using masking and spoofing techniques.

All types of bots ($NA$, $A$, and $A^*$) are capable of injecting audio (such as synthesized/recorded voice) into the 911 call. Doing so can be done to incur extra delays in the 911 call taker's process of determining whether the caller is a bot or not.

## 5. ANATOMY OF A 911-DDOS BOT

In order to demonstrate the feasibility a 911-DDoS attack launched from anonymized mobile botnet, we designed, implemented and tested versions of the $A^*$ bot on a small cellular network. For comparison purposes, we also implement $A$ and $NA$ type bots.

### 5.1 Implementation of an $NA$ Bot

Non-anonymized DDoS attacks on 911 can be launched from the main OS without interfering with the baseband firmware. Although this is not the main contribution of the paper, we now briefly present the implementation of such a bot for comparison purposes.

For our tests, we implemented a $NA$ type bot on the Android OS. In the following subsections we briefly describe the Android Radio Interface Layer (RIL), the bot's RIL interception proxy, and the implementation for injecting audio into emergency calls.

#### 5.1.1 Radio Interface Layer (RIL)

In Android, the Radio Interface Layer (RIL) is the interface between the high level telephony services and the baseband hardware (cellular modem). The RIL defines two types of operations: 1) solicited commands (such as calling requests and the sending SMS messages) are sent from the telephony services to the baseband, and 2) unsolicited commands (such as incoming calls and network notifications) are sent from the baseband to the telephony services. Each vendor supplies its own implementation for the RIL interface. A vendor's RIL is closed source and shipped with the stock Android firmware as shared object binary files. The RIL Daemon (RILD) serves as the interface between the Android telephony services and the vendor's RIL via a communication socket.

#### 5.1.2 The RIL Interception Proxy

To demonstrate a $NA$ type bot's covert abilities, we implemented an interception proxy mediating between the RILD and the vendor's RIL. This component is capable of intercepting and manipulating any solicited or unsolicited command exchanged between the Android framework and the baseband.

#### 5.1.3 Making a Covert Emergency Call

The following are the primary commands which are handled by the RIL interception proxy to perform its DoS activities under stealth. The specific commands may differ between manufacturers. When the user makes/receives a call, the bot immediately halts its activities. With REQUEST_DIAL command the bot initiates an emergency call at its interception proxy towards the vendor's RIL. After initiating the emergency call, the proxy intercepts the response as it comes up from the baseband. The Android framework is not aware that the call has been made or received, because the commands did not pass through the standard framework's APIs. The CALL_STATE_CHANGED command is sent from the baseband to the Android telephony services whenever the emergency call state changes (e.g., when an emergency call is answered). The RIL intercepts this command to hide the bot's activity. The RIL_REQUEST_GET_CURRENT_CALLS command is used to get the list of current calls and their status. When the Android telephony service invokes this command it notifies the respective elements (e.g., the GUI) and updates the logs/call history. The bot's proxy intercepts this list, omits the emergency call, and forwards the list up to the RILD.

#### 5.1.4 Audio Injection

By default, the injection of audio into an ongoing call (uplink) is not supported by the Android API. On some platforms (e.g., Qualcomm's MSM8960 chipset) it is possible to inject recorded audio, utilizing support from the Advanced Linux Sound Architecture (ASLA) via the *Incall_Music* Audio Mixer. Using the technique described in [31], we were able to successfully inject a WAV audio recording into an ongoing call using a Samsung Galaxy S4 Mini smartphone.

### 5.2 Implementations of $A$ and $A^*$ Bots

Modern mobile phones have a cellular protocol stack managed by a separate real-time operating system (RTOS) which runs on the baseband processor. The main OS (e.g., Android) and the baseband RTOS work independently of one another and exchange data through shared memory or similar mechanisms.

In order to make anonymous 911 calls, $A$ type bots need to switch the protocol stack into no-SIM state, and $A^*$ bots need to be able to spoof the phone's identifiers. These manipulations require low-level access to the phone's cellular protocol stack and internal state machine.

To demonstrate these bots, we had to implement a rookit within the baseband firmware, which can modify the baseband's RTOS and manage its state machine. There have been several cases where attackers have installed malware within a device's firmware in order to perform malicious [32] [33]. In particular, baseband malware exploits and attacks are discussed and demonstrated [34] [35] [36] [37].

#### 5.2.1 OsmocomBB

To make changes to a phone's protocol stack, we used a GSM baseband software called OsmocomBB [38]. OsmocomBB is the only way to freely research the implementation of a mobile's baseband software. OsmocomBB has been used for implementation in academic research and other publications. The project currently supports several mobile phone models, of which we used the Motorola C123. Note that the implementation concepts in this section are relevant to UMTS and LTE as well [39].

#### 5.2.2 Protocol Stack Modification

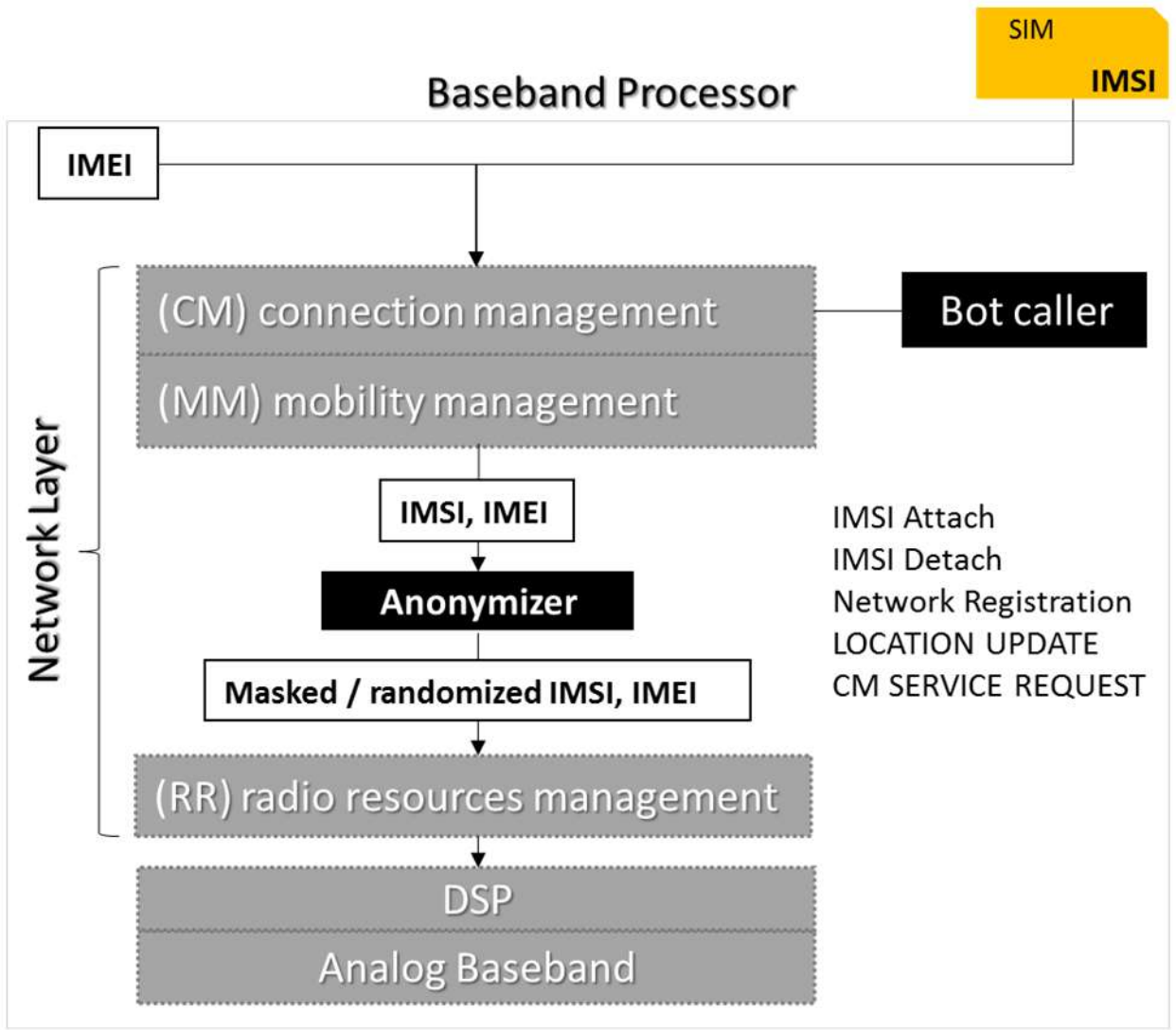

**Figure 3. The architecture of the 911 DDoS bot within the firmware of the baseband processor.**

The protocol stack in OsmocomBB consists of three main layers. From bottom to top they are: L1 - the radio interface,

L2 - the data link layer, and L3 - the network layer. L3 - is divided into three sublayers, and from bottom up they are: L3a - radio resources management (RR), L3b - mobility management (MM), and L3c - connection management (CM). Based on the OsmocomBB 'mobile' application, we focus our implementation on the L3b and L3c sublayers where the network registration procedures are handled. By default, OsmocomBB has its device transmission (Tx) capabilities disabled. Therefore we had to compile the firmware with the DCONFIG_TX_ENABLE flag in order to enable outgoing calls.

The main architecture of the bot is depicted in Fig 3. Our bot is implemented as a rootkit within the firmware of the baseband processor and is compound of two components (1) a bot caller, and (2) an anonymizer. The bot caller component manages the 911 call initiations (starting and stopping the DDoS attack). It handles the state machine, and enter and leave to/from anonymous mode. The call initiation is done by building emergency call requests via the CM and MM layers in the protocol stack. The anonymizer component resides between the MM and RR layers. It is aimed at hiding the device's identity by masking or randomizing the device's IMSI and IMEI identifiers before they reach the radio layer.

### 5.2.3 *Anonymized (A bot)*

When a mobile phone is switched on, its baseband processor starts an IMSI attach procedure with the cellular network. During the IMSI attach, the device sends a "location update request" message which include its IMSI or TMSI. The Home Location Register (HLR) validates the IMSI by ensuring that it is known to the network and is allowed subscriber services. After the IMSI validation, the network sends an "authentication request" message back to the mobile phone, which answers with a calculated key (SRES, a 32-bit Signed Response). Once authenticated by the network, the mobile phone receives a "location updating accept" message which completes its registration process. To unregister from the network, the mobile phone sends an "IMSI detach" message which includes its IMSI.

In no-SIM state, a mobile phone behaves like a device without a SIM card, and therefore lacks an IMSI. In GSM, UMTS and LTE standards, the device will not attempt to perform location updates to the network and rejects any request for MM connections except for making emergency calls [40]. Consequentially, in no-SIM state the device remains unidentified to the cellular network, yet can issue emergency calls.

In order to put the device into no-SIM state, we patched the SIM initialization MM related procedures in L3. The anonymizer initiates a baseband report that the device SIM card is missing (in src/target/firmware/calypso). In order to disconnect the mobile phone from the network, the anonymizer invokes MMR_NREG_REQ [41] and initiates an IMSI detach procedure with the network. At the end of this process, the device is 1) in no-SIM state, 2) no longer registered to the network and, 3) able to initiate emergency calls as an NSI device. The emergency calls are issued from our bot caller component using CM SERVICE REQUEST with a service type information element set to emergency call establishment (0010).

### 5.2.4 *Persistent Anonymized ($\boldsymbol{A}^*$ bot)*

Persistent anonymized bots provide the highest degree of anonymization in a cellular network. They continuously spoof the IMSI or IMEI in order to evade blacklisting and identification systems. We present two ways of implementing $A^*$ bots: IMEI/IMSI Masking and IMSI/IMEI Spoofing.

**IMEI/IMSI Masking.** The latest cellular standards require that a mobile phone must use its IMEI for identification in emergency calls (CM SERVICE REQUEST) when no IMSI is available during the IMSI attach procedure [24]**.** In no-SIM state the IMSI is absent, hence basebands which are compliant with the standards must use the IMEI as an identifier to the network. $A^*$ bots have a higher degree of anonymity, since their identifiers cannot be traced back to the device. They are also persistent in that they cannot be blacklisted, since they are don't expose any fixed identifier to the network. The process is the same as we proposed for the $A$ type bot (no-SIM state) but with the addition of IMEI spoofing. For each initiation of an emergency call, we override the device's current IMEI and replace it with a random, yet valid, IMEI value. As a result, different IMEI identifiers will be exposed to the network during sequential emergency calls.

The outline of this variant is presented in Algorithm 1. Initially, the anonymizer invokes an IMSI detach procedure, and the device enters the no-SIM state (lines 3-4). The actual DDoS is accomplished within the bot caller, by repeatedly calling 911 and holding each call until a call-end event (lines 6-10), each time with a random valid IMEI (line 7). After a specified number of calls, or when the user presses the keypad to initiate a call, the device returns to its normal functionality (lines 12-14).

**Algorithm 1** $A^*$ Bot (No-SIM + IMEI Spoofing)

```
1:  procedure Start-DDoS
2:      // set the device state
3:      Invoke(IMSI-Detach-procedure)
4:      SetDeviceState(No-SIM)
5:          // main DDoS loop. Stops when user activity detected
6:      while (no key-pressed) do
7:          SetIMEI(randomIMEI)
8:          InitiateEmergencyCall()
9:          yield(Call-End-event)
10:     end while
11:     // restore the device state and attach
12:     SetIMEI(origionalIMEI)
13:     SetDeviceState(Attached-SIM)
14:   Invoke(IMSI-Attach-procedure)
15:   Return
```

```
Algorithm 2   A* Bot (IMSI + IMEI Spoofing)
1:  procedure Start-DDoS
2:   // main DDoS loop. Stops when user activity detected
3:      while (no key-pressed) do
4:         // set random IMSI and attach
5:         Invoke(IMSI-Detach-procedure)
6          SetIMSI(randomIMSI)
7:         Invoke(IMSI-Attach-procedure)
8:         InitiateEmergencyCall()
9:         // Call N times with the same identity
10:        for (i=1,...,N) do
1:             InitiateEmergencyCall()
12:            yield(Call-End-event)
11         end for
10:     end while
11:    // restore the device identifiers and attach
12     Invoke(IMSI-Detach-procedure)
13:    SetIMSI(originalIMSI)
14:   Invoke(IMSI-Attach-procedure)
15:   Return
```

**IMEI/IMSI Spoofing.** In this variant of $A^*$, the mobile phone behaves like it has a SIM card, but it supplies the network with a spoofed IMSI. As a result, the device-network registration fails with a LOCATION UPDATING REJECT. However, the device is still permitted to make emergency calls as NSI device. This means that during the IMSI attach procedure the location update request is rejected by the network, since the IMSI is unknown. The network sends a LOCATION UPDATING REJECT message and the *reject cause information element* is set to the IMSI_UNKNOWN_TO_HLR or IMSI_UNKNOWN_TO_VLR values. In this state the device functionality is limited to emergency calls.

We implement this variant by randomizing the IMSI of the device during the device's location update procedure gsm48_mm_loc_upd_normal(). In this way the device registers with the network using a different IMSI after every few call, enabling it to attack under a new unknown identity. The outline of the IMSI spoofing operational mode is presented in Algorithm 2.

Initially, the anonymizer performs the IMSI attach procedure with a randomized IMSI (line 1). The actual DDoS is accomplished by repeating emergency calls (lines 2-6) and subsequently invoking the IMSI detach procedure. At the end of the attack, an IMSI attach procedure with the real IMSI is invoked.

The difference between the masking and spoofing variants of A* is that wireless carriers may be less able to block bots implementing IMSI spoofing. Recall that IMSI spoofing causes a NSI state of unknown IMSI. By blocking any unknown IMSI, legitimate roaming users would be blocked from 911 services, which is forbidden by the FCC regulations. On the other hand, the IMSI spoofing variant is slower than the IMSI masking variant, since it invokes the network IMSI attach and IMSI detach procedures frequently.

Table 3 summarizes this information and presents a comparison between the different types of 911 bots and their characteristics.

**Table 3. Type of implemented bots.**

| # | *Technique* | *Bot type* | *Implementation* | *Comments* |
|---|---|---|---|---|
| **1** | No-SIM | $\boldsymbol{A}$ | Baseband rootkit | Simple to implement |
| **2** | No-SIM, IMEI Spoof | $\boldsymbol{A^*}$ | Baseband rootkit | Fastest operation |
| **3** | IMSI & IMEI Spoof | $\boldsymbol{A^*}$ | Baseband rootkit | Highest anonymity |
| **4** | IMSI, IMEI exposed | $\boldsymbol{NA}$ | Android OS Malware | OS-Level bot |

### 5.3 Test Environment

In order to test and evaluate the implemented bots, we built a small cellular network in our lab (Fig. 4). This testbed allowed us to directly examine the traffic between a mobile phone and a cellular network. In particular, it enabled us to examine the transfer of device identifiers, network registration procedures, IMSI attach and IMSI detach procedures, and emergency call setups –all from the perspective of the cellular network.

For the base transceiver station (BTS) we used the *ip.access* nanoBTS hardware [42] (Fig. 4a). This is a small BTS with an Abis interface built in accordance with ETSI standards, thereby guaranteeing compatibility with existing handsets.

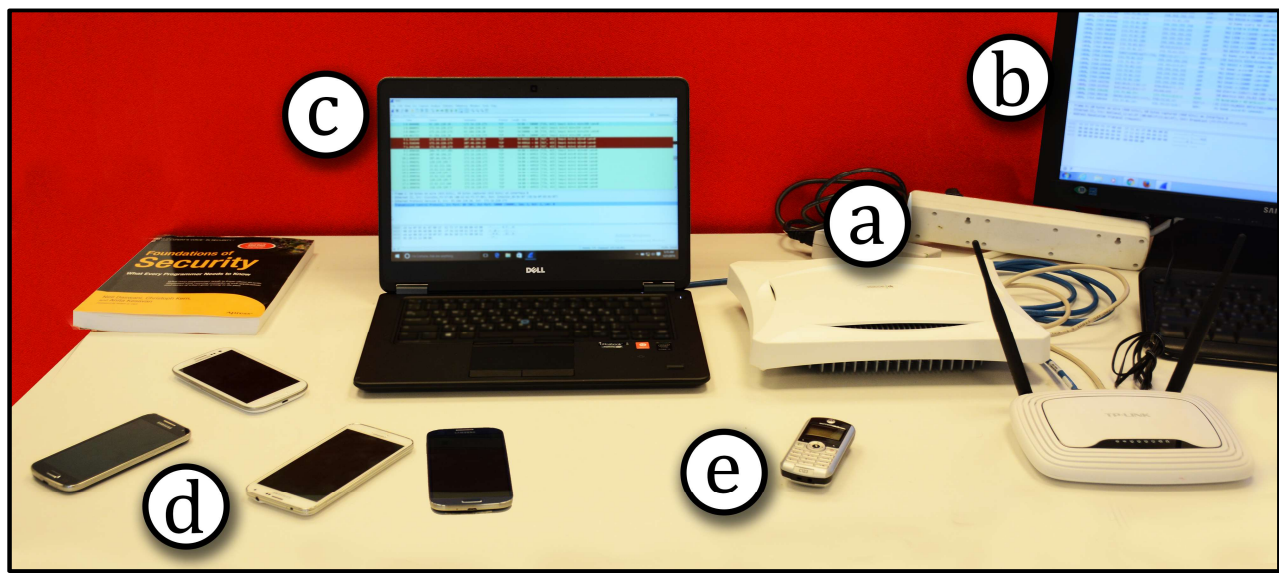

**Figure 4. The testing environment in the lab.**

We used the GPRS/GSM 900 model (165CU) which supports telephony services, speech codecs, and GPRS. The traffic to/from the nanoBTS was captured on an external desktop computer (Fig. 4b) via an Ethernet connection using the Wireshark and TShark tools. The nanoBTS was connected to a laptop (Fig. 4c) which simulated the network's BSC, MSC, Home Location Register (HLR), and Visitor Location Register (VLR) using openBSC [43].

With this testbed we successfully verified the $NA$, $A$, and $A^*$ bot concepts. The $NA$ type bot was tested on the Samsung Galaxy S3, S4, and S5 smartphones running the Android 4.4 (KitKat) and Android 5.x (Lollipop) OSs (Fig. 4d). In-call audio injection was verified on a Samsung Galaxy S4 Mini smartphone. The $A$ and $A^*$ type bots required a modified protocol stack and were tested on a Motorola C123 mobile phone with a Calypso baseband (Fig. 4e). During the tests we checked the no-SIM and IMEI/IMSI spoofing operational states. We monitored the IMSI and IMEI identifiers within the relevant network registration, location updates, and call procedures.

## 6. SIMULATION RESULTS & ANALYSIS

In this section we verify whether anonymous unblockable 911 bots ($A^*$) are a significant threat. We accomplish this by (1) modeling the current E911 infrastructure based on surveys and gathered statistics, (2) examining the model to DDoS vulnerabilities in its structure, and (3) quantifying the impact of the attack by simulating attacks over the model.

### 6.1 The E911 Network Model & Topology

We model the E911 network as the bidirectional multigraph $G \coloneqq (V, E)$. Each vertex in $V$ has an associated type $T = \{\text{PSAP}, \text{SR}, \text{CO}, \text{MSC}\}$. An edge between two vertices in $G$ has a multiplicity equal to the number of trunks between them. A trunk can carry exactly one voice call at a time. $G$ has the following topological constraints: vertices of type CO, MSC, and PSAP can *only* be connected to vertices of type SR. Furthermore, vertices of type SR *may* be connected to other vertices of type SR (referred to as tandem connections). Lastly, vertices of type CO, MSC, and PSAP *may* be connected to more than one vertex of type SR and vice versa. As a visual reference, please refer to Fig. 1 in section 3.

Let $\Gamma_t(v)$ be the set of vertices of type $t \in T$ which have an edge shared with vertex $v$. For example, $\Gamma_{\text{PSAP}}(s_i)$ is the set of PSAPs connected to the SR $s_i$. It is important to note that the network elements of $\Gamma_{\text{PSAP}}(s_i)$ are all located in the same geographical region as $s_i$.

Given the assumed model above, we surveyed published reports online and produces three instances which reflect the US's current infrastructure at city, state, and country levels:

**Country-level.** In December 2015, there were 7,227 active PSAPs listed in the FCC master PSAP registry [44]. Explicit information on E911 networks is generally not available to the public. Therefore, we model the entire US E911 network based on general statistics and reports published from 2014 to 2015 [45]. We use linear regression to complete the missing wireline/wireless call volume statistics for 27 states. In Fig. 5 we present each state's call volume composition, where the red line indicates the average percent of wireless calls across the country (72.8%).

Since information about all of the SRs in the US is unavailable, for each state in the country-level analysis we generalize the topology by assigning one SR to that state's PSAPs. In order to determine each individual PSAP's call volume, we divided the state's call volume across its PSAPs scaled according to the PSAP's local population [46]. In Fig. 6 we plot the locations of every primary PSAP scaled according to their call volume. We assume that the PSAP's number of consoles, trunks, and trunk types, reflect the configurations of NC's PSAPs from [47] as presented in Table 4.

**State-level.** For a more detailed analysis, we model the state of North Carolina's (NC) E911 infrastructure. In 2008, a survey detailing the state of NC's E911 topology and network statistics was published [47]. In 2008, the state of North Carolina had a population of approximately 9.2 million and was the tenth largest state in the US. NC had 20 SRs managed by AT&T, Verizon, Embarq, Citizens Telephone, and Windstream, and 188 PSAPs with a total of 775 call taker positions that handled an annual 911 call volume of 8,412,700 (23,048 calls daily).

Using the data in [47] we were able to reconstruct most of NC's E911 network topology. We refer to the reconstructed topology as $G_{NC}$. Fig. 7 visualizes $G_{NC}$, where red nodes are SRs, and the PSAPs are color coded according to their respective community. To determine each SR's total *inbound* traffic volume, we aggregated the call traffic from the PSAPs to the SRs. More formally, given the SR $s$, $vol(s) = \sum_i^{|\Gamma_{\text{PSAP}}(s)|} vol(p_i)$ where $p_i \in \Gamma_{\text{PSAP}}(s)$ and $vol(v)$ is the average daily inbound call volume of vertex $v$.

In comparison with the FCC's master PSAP registry [44] we found that only one new PSAP (out of 188) has been added to NC between 2008 and 2016. Therefore we believe the information in [47] can be used to gain relevant insight into the current E911 infrastructure of US cities and states, and we select NC's E911 network for our in depth analysis.

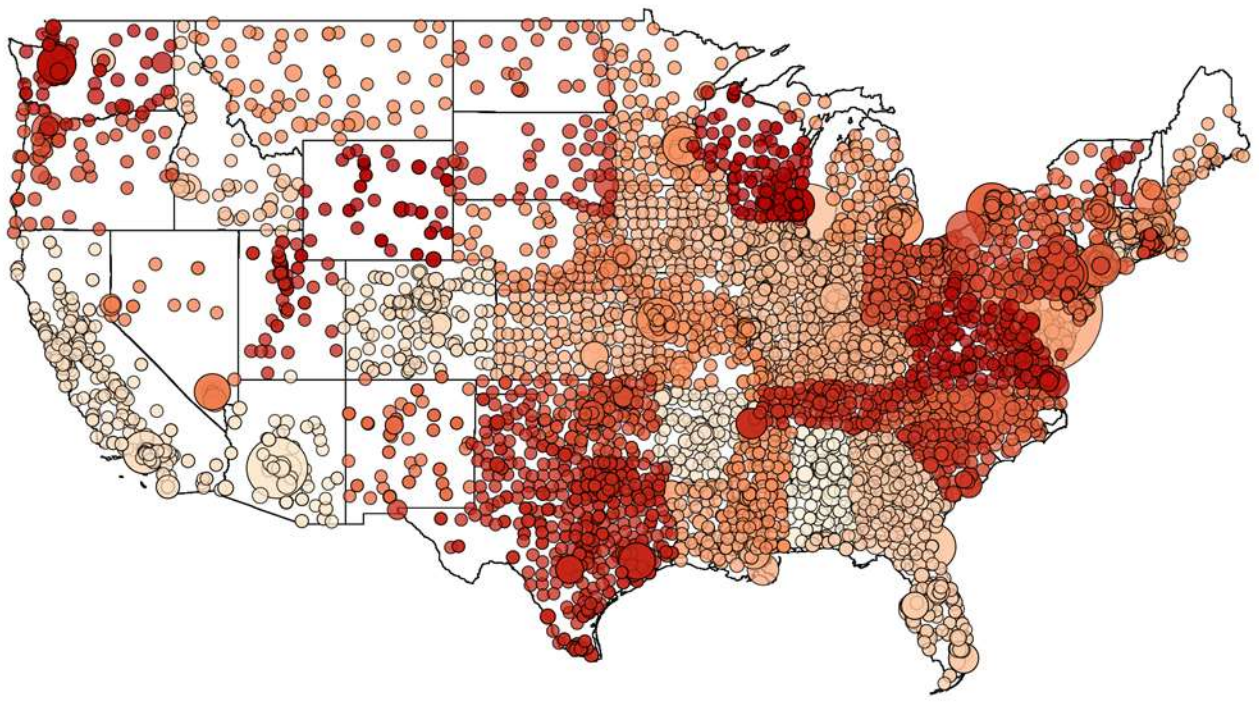

**Figure 6. The locations and call volumes of the primary PSAPs on the US mainland, where the shades correspond to different states.**

| Configuration per PSAP | Setting |
|---|---|
| Number of consoles per 10K population | 1.1925 |
| Number of Trunks per 10K population | 1.7053 |
| Ratio of *wireline / wireless / both* trunks types | 0.1602, 0.0951, 0.7448 |
| Percent of PSAPs that have no dedicated trunks | 60% |

**Table 4. Configurations applied to each PSAP (country-level).**

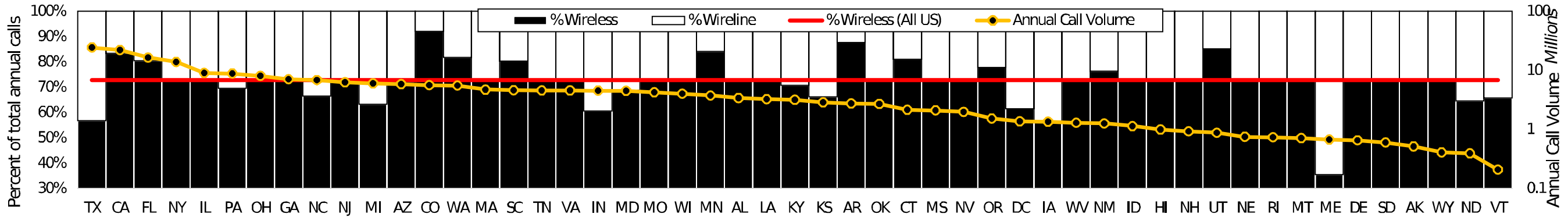

Figure 5. The percentage of each state's wireline and wireless call volumes, sorted by highest annual call volume (left) to lowest (right).

**City-Level.** We select the city of Charlotte for our detailed analysis since it is the largest city in North Carolina with a population of nearly one million. The survey in [47] shows that the city had three PSAPs which handled a 911 call volume of 1,014,056 annually (2,778 calls daily). The survey also shows that the city's PSAPs are served by three SRs serviced by AT&T. These SRs support an *additional* 25 PSAPs outside of the city. The traffic of these external PSAPs directly affects the load on the SRs serving the city. We denote the subgraph containing the 25 PSAPs, the 3 PSAPs serving Charlotte, and all contained edges as Charlotte's PSAP community $G_{Cha}$ , where $G_{Char} \subset G_{NC}$.

## 6.2 Topological Vulnerability Analysis

**Network Robustness.** NC's 911 network has a heavy-tail degree distribution. These types of networks are robust to random failures (random vertex removals) but highly intolerant of targeted attacks (i.e., removal of the nodes with the largest degree). This means that should a central SR, such as Rocky Mount, get overloaded with DDoS calls, the 911 system will be severely disabled. Consequently, should a DDoS attack disable the Rocky Mount SR, 64 PSAPs (~31% of the population) would be without access to 911 services. To make the system more robust, it is important that COs and MSCs be connected to redundant SRs. The NC's PSAPs SR map is given in Appendix A.

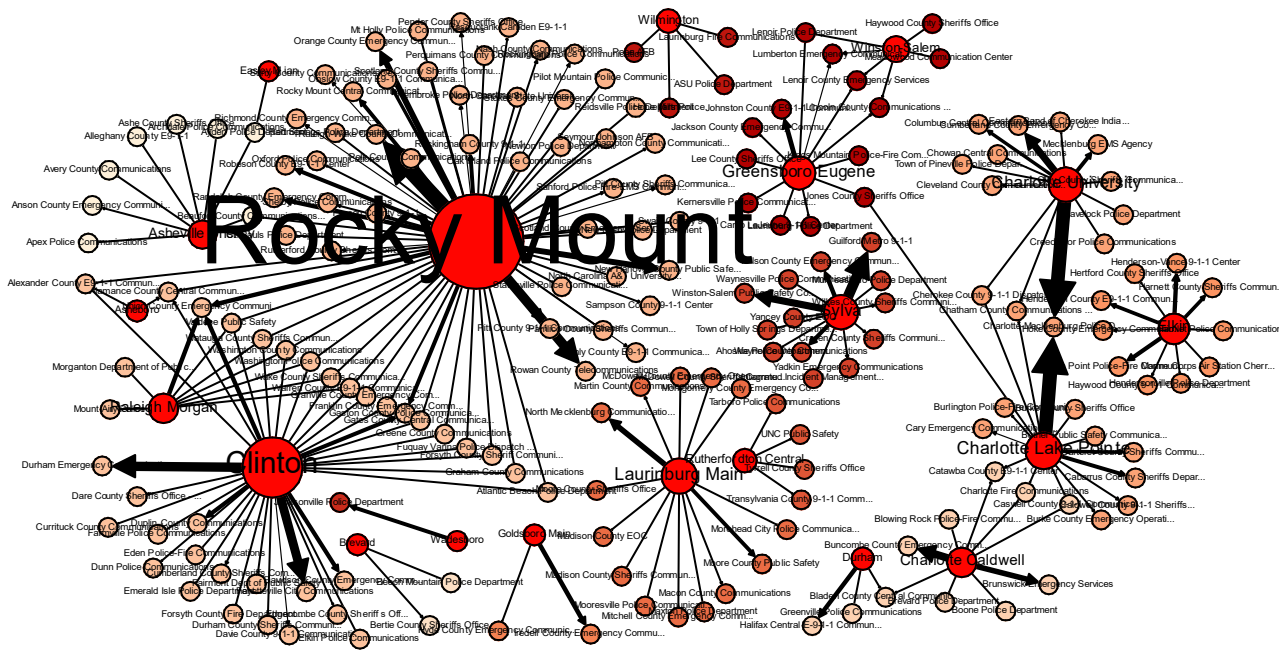

**Figure 7. The E911 network topology of North Carolina.**

**Selective Router Redundancy**. Fig. 7 shows some regions have SRs which can act as redundancies encase of a traffic overflow. However, 83% of NC's PSAPs are connected to a single SR. For example, Charlotte's three PSAPs are only supported by the *Charlotte Caldwell SR*. The existing redundancies do not reflect the underlying population size. This puts more people at risk in the event of a DDoS attack due to single node failures.

**Traffic Overflows.** We note that a DDoS attack launched from one geographic region can have an effect on neighboring regions. For example, if the city of Statesville undergoes a localized DDoS attack, the call volume will overwhelm the serving SRs (*Caldwell* and *Lake Pointe*). This will result in an outage of 20-22% of NC's 911 services (measured by call volume). The overflow of legitimate calls in the *Caldwell* and *Lake Pointe* regions may be transferred to the *University SR* and overwhelm that SR as well. Note, SRs are sometimes programmed with overflow routes to default PSAPs regardless of the PSAP assigned to the caller in the SRDB [48]. Therefore, if the *University SR* is not overloaded, then some of its PSAPs may be overloaded with the extra traffic.

**PSAP Workload.** NENA requires that PSAPs provide a grade of service (GoS) of 0.01 (i.e., no more than one out of every 100 calls are unanswered) [49]. From the data in [47] we calculated that 77% of the NC PSAPs believed that they needed a 22.3% and 41% increase of full and part-time call takers, respectively. Charlotte's large PSAP (PSAP 2) reports that it needs an extra 33 full-time call takers, although it is only budgeted for three more. Therefore, the call taker workforce of NC is understaffed and may have difficulty meeting NENA's requirement. This staffing problem is a well-known issue across the US and causes legitimate 911 calls to get "the busy signal" on a regular basis [50]. This reality makes the threat of even a *moderately* sized DDoS attack much more significant. Moreover, we calculate that 86% of all call taker positions in NC hold the dispatcher responsibility as well (i.e., the one who contacts the fire, police, or medical services). This affects the rate at which calls can be processed.

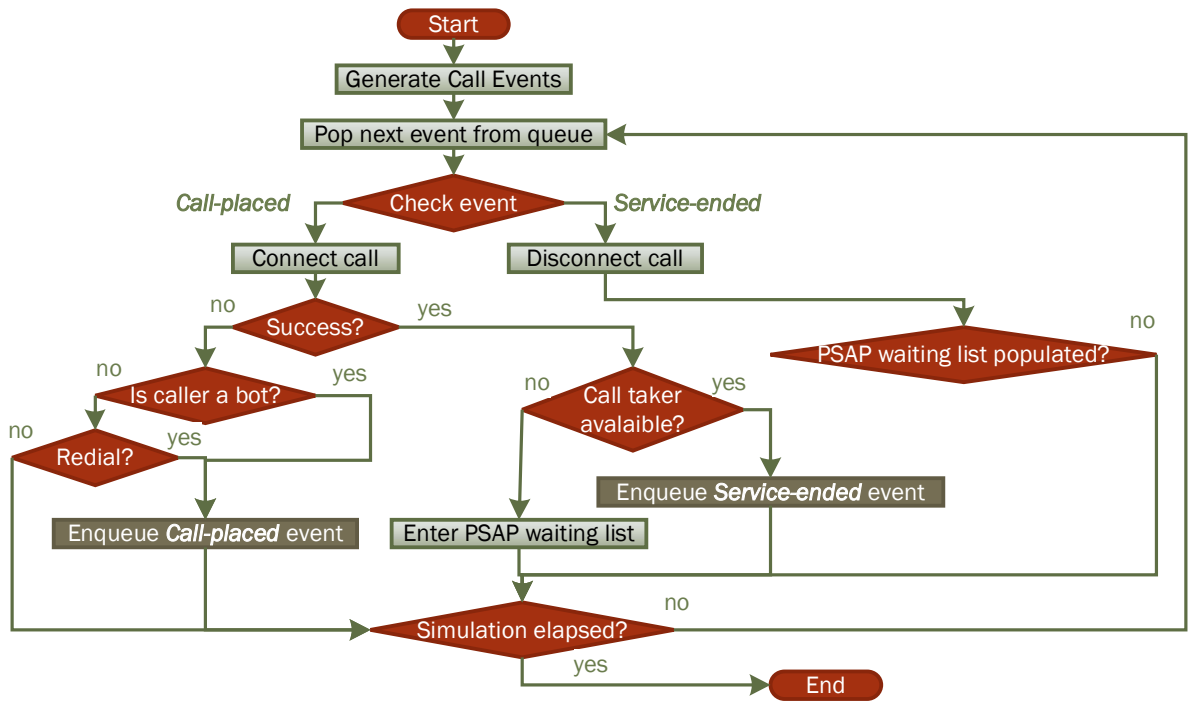

**Figure 8. A flow chart describing the operation of the DES.**

**PSAP Call Balancing.** When a voice circuit is established between two endpoints in telephony networks, the respective trunks between each of the involved switches are occupied by the call. Therefore, when a SR connects a call to a PSAP, the trunk used between the SR and the PSAP is busy until the call is ended by the caller or the PSAP. Charlotte does not have any trunks dedicated for wireline calls. This means that the traffic from a mobile DDoS attack can block out *all* legitimate 911 callers, even if the trunks are occupied for only a few seconds at a time.

**PSAP Call Waiting.** If there are more trunks than call taker positions, callers can be put on hold until they can be answered. More formally, let $tr(p)$ and $c(p)$ be the number of PSAP $p$'s inbound trunks and call taker (console) positions. We calculate the queue length of $p$ as,

$$Q(P) = \begin{cases} tr(P) - c(P), tr(P) \geq c(P) \\ 0, else \end{cases} \quad (1)$$

Using (1) on the data in [50] we found that only 67% of NC's PSAPs have queues where the average queue length is 1.4. Charlotte's PSAPs have no queues at all. A PSAP with no queue blocks incoming calls while all its call takers are busy. These delays have immediate repercussions to the caller's personal safety, because in some emergencies, every second counts.

## 6.3 Attack Simulation & Impact

### *6.3.1 The Discreet Event Simulator*

In order to measure the impact of a mobile phone DDoS attack on the selected E911 networks, we use a discreet event simulator (DES). The simulation input parameters are the simulation duration ($t_{simDur}$), DDoS start time ($t_{ddosStart}$), and number of bots ($n_{bot}$). The flow chart in Fig. 8 summarizes how the DES works. The following are the base assumptions taken for the simulation.

**Time of day.** Since PSAPs are generally fully staffed during the busy hour of the day, we simulate the DDoS at this time (the best-case scenario). We assume that the busy hour of a PSAP contains 15% of the daily call volume (denoted as $\alpha_{bh}$) as used in the official NENA staffing guide [51].

**Call Event Generation.** In our DES, calls enter the simulation at the intake of the network's SRs. Let $\lambda_s = \frac{vol(s)\cdot\alpha_{bh}}{60*60}$ be the average call arrival rate of legitimate calls during the busy hour for the SR $s$ measured in calls per second. For each $s_i \in S$ we randomly generate that SR's legitimate calls over a Poisson distribution using $\lambda_{s_i}$. According to the report in [25], 70% of all 911 calls are wireless. We assume that the call types are distributed uniformly across the population, so that for a randomly selected caller $p(\text{wireless}) = 0.7$ and $p(\text{wireline}) = 1 - p(\text{wireless})$. The bots are distributed at the SRs in proportional to the SRs' legitimate call volume.

**Call Processing Time.** We model the legitimate call service time of PSAP $p$ with an exponential distribution, where $\lambda_p$ is provided in [47] for the state-level, and for the country-level we take 60 and 90 seconds for wireline and wireless calls, respectively [51]. We assume that the call service time for a bot is approximately six seconds: the time it takes for a call taker to say "9-1-1, what is your emergency?" and listen for a few seconds to be sure that the call is indeed a bot.

**Redials.** Let $p(\text{recall})$ be the probability of a legitimate caller redialing 9-1-1 after being blocked. In a real emergency, a legitimate caller will persist in calling 911 until his/her call is answered. Therefore, we set $p(\text{recall})$ to 0.85 as used in [52]. We assume that the redial delay has an exponential distribution with an average of $\beta = 20$ seconds (with a four second overhead for the end-to-end call setup time [53]).

**SR Capacity.** Since a SR is a telephony switch, it can be expected that the SR $s$ has a capacity similar in magnitude to $\sum tr(\Gamma_{\text{PSAP}}(s))$. However, in the simulation we do not limit the SR's capacity but rather focus on the PSAP's capacity instead. Later we will analyze the SR load.

### *6.3.2 Simulation Results*

**Caller Experience.** At the state-level, we found that as little as 6,000 bots (0.0006% of NC's population) is enough to deny 20% and 50% of wireline and wireless callers from ever reaching 911 services (after 4-5 attempts each per caller). This is even more significant considering that 70% of 911 calls are wireless. With 50K bots (0.0054% of NC's population) nearly 90% of all wireless 911 callers never reach a call taker. Fig. 9 presents the percent of callers who give up on 911 services for different numbers of bots attacking NC.

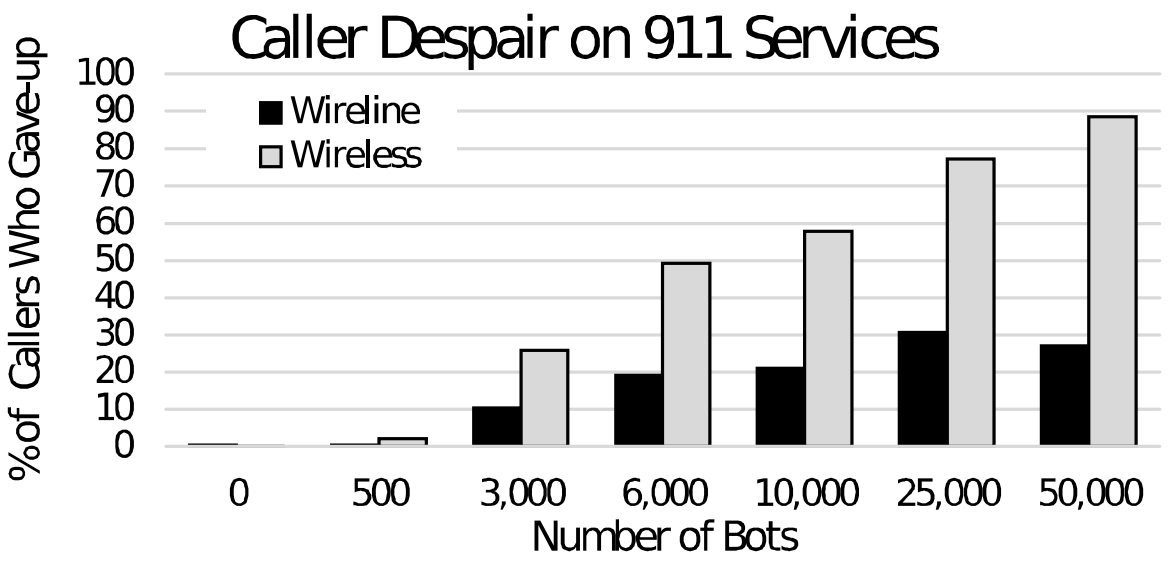

**Figure 9. The percent of NC's 911 callers who give up on reaching 911 services under different sized DDoS attacks.**

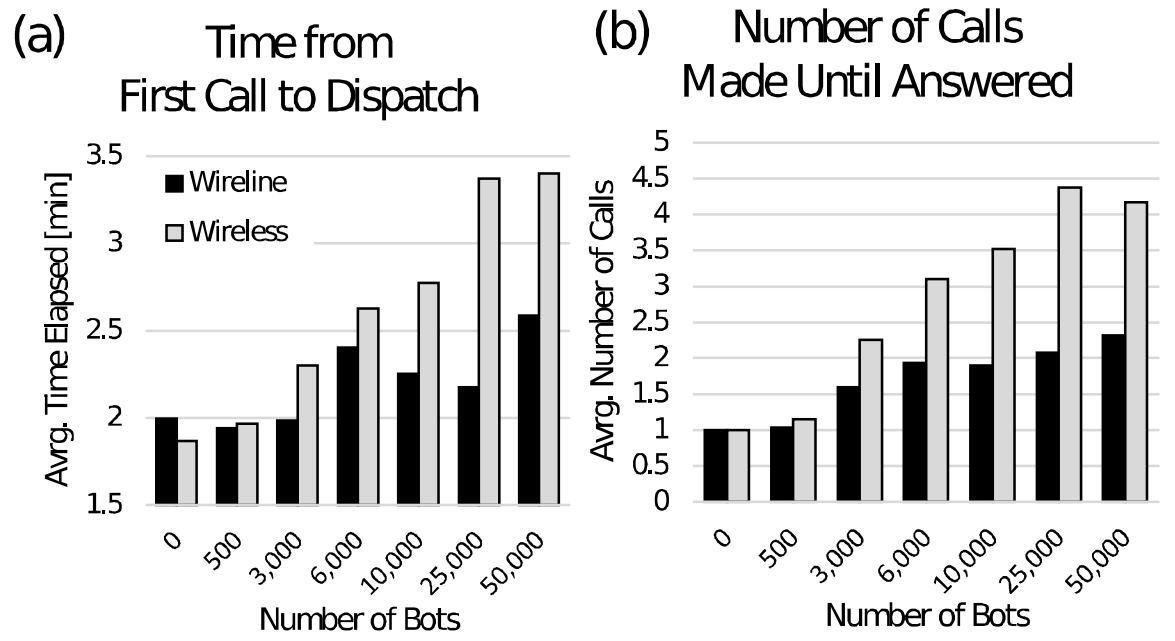

**Figure 10. The effect of the number of bots on NC's 911 call response time**.

| | *Percent of Callers who Give up on Reaching 911* | | | *Avrg. Number of Calls Made until Answered* | | |
|---|---|---|---|---|---|---|
| *Number of* | **Wireline** | **Wireless** | **All** | **Wireline** | **Wireless** | **All** |
| **0** | 0.13% | 0.11% | 0.11% | 1 | 1 | 1 |
| **100,000** | 8.87% | 15.30% | 13.56% | 1.32 | 1.56 | 1.49 |
| **200,000** | 23.06% | 37.16% | 33.36% | 1.85 | 2.59 | 2.36 |
| **320,000** | 34.04% | 51.88% | 47.08% | 2.02 | 3.24 | 2.83 |
| **500,000** | 40.27% | 63.61% | 57.32% | 2.11 | 3.91 | 3.23 |
| **800,000** | 48.59% | 74.46% | 67.49% | 1.96 | 4.3 | 3.3 |

**Table 5. The effect various sized attacks on the entire US have on legitimate 911 callers.**

Since bots make wireless calls, they do not impact the wireline caller's give up rate after 25k bots. This is because some PSAPs have dedicated wireline trunks. However, the city of Charlotte's PSAPs have no dedicated trunks. As a result, 6K bots in NC (2.8K falling in the jurisdiction of Charlotte's PSAP community) is enough to deny ~80% of Charlotte's 911 callers from emergency services.

For those in NC who successfully reach a call taker, there is a significant amount of time wasted in redial attempts. Fig. 10 shows (a) the average service time (time elapsed between the first call and dispatch) and (b) the average number of calls made until answered for different sizes of DDoS attacks. Fig. 10 shows that with 6K bots, increases the service time by 40%. In this scenario, a caller would wait an *additional* 45sec-3min (factoring one standard deviation) and call an average of three times to get emergency service.

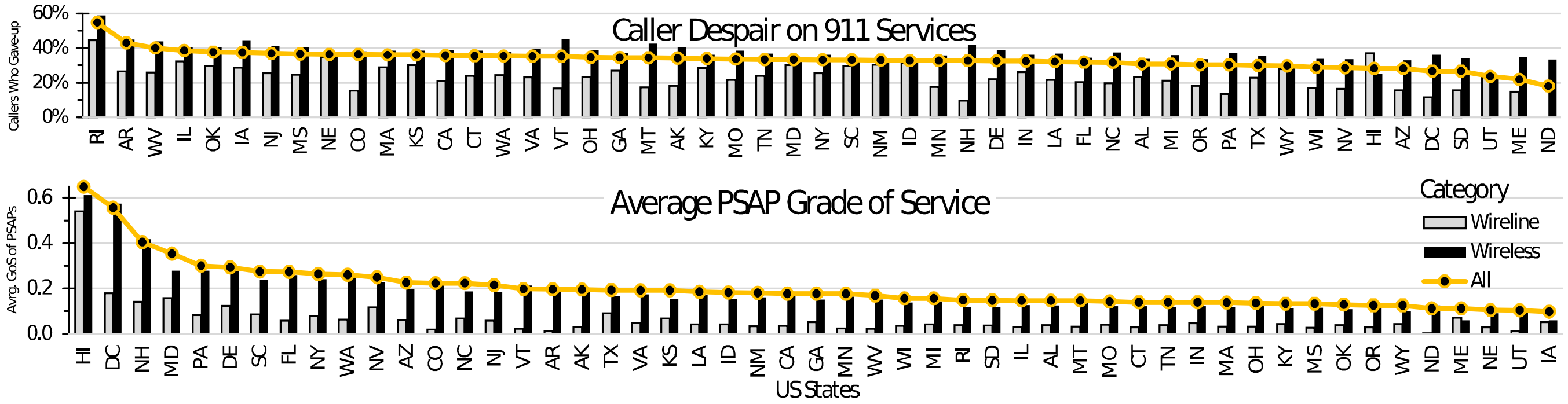

**Figure 11. The impact of 200K bots across the US has on each state's (top) number of caller despairs and (bottom) average PSAP GoS.**

At the country-level, we found that as little as 200,000 bots, distributed across the population of the US, is enough to significantly disrupt 911 services across the US (Table 5). This means that an attacker only needs to infect ~0.0006% of the country's population in order to successfully DDoS emergency services (In 2014, the USA has a population of ~319 million [46]). Under these circumstances, an attacker can cause 33% of the nations' legitimate callers to give up in reaching 911.

**PSAP Performance.** Some regions of NC receive a lower GoS than others because not every PSAP is configured the same. For instance, NC has 28 PSAPs that have 2 call taker positions, and daily call volumes ranging from 1 to 58 calls. In light of this reality, it is clear that a caller's accessibility to 911 during a DDoS attack is dependent on his/her location (i.e. serving PSAP). On average, during a 6K bot attack, the percent of NC residents which are denied service (given a busy-tone) is 75% and 91% when calling from wireline and wireless devices respectively.

The bottom of Fig. 11 shows the average for each state of the US. Note that a state's average PSAP GoS does not reflect the number of callers who give up in that state. This is because the GoS also takes into account calls which were rejected by a PSAP, but handed over to another one. Fig. 11 shows that all states exceed NENA's requirement of a GoS of p0.01 [49] when under a relatively small attack.

**Network Performance.** An overload in the network itself prevents 911 calls from reaching PSAPs. Legitimate callers increase the call volume during an attack because they redial if blocked. We found that a 6K botnet attack can cause a 160% increase in NC's *legitimate* call volume.

Calls which are not connected to a PSAP (bot and legitimate) add to the SR's load as well. This is because the TS-SR trunk is occupied for at least the time it takes to setup the trunk between the switches. This call setup time is 3.9 seconds when using multi-frequency (MF) signaling and 100ms when using the SS7 network [54]. Although MF is an old signaling method, it is still may be employed in some areas. We will assume that $\sum tr(\Gamma_{MSC,CO}(s)) = \sum tr(\Gamma_{PSAP}(s))$. The offered load of a single trunk (one voice line), when measured in Erlangs, is the number of traffic hours offered to the trunk per hour. Therefore, a SR is overloaded if it has an offered load greater than an average of one Erlang per trunk.

With SS7 signaling, the DES showed that NC will have 37% of its SRs overloaded during a 6K sized attack affecting 45% of NC's 911 callers. Moreover, for old networks employing MF signaling, 87% of SRs would be overloaded affecting 98% of 911 callers. For this reason SRs which still employ MF signaling should be immediately considered for a SS7 signaling upgrade.

In summary, the results from this section show that only a few anonymous 911 bots ($A$) are enough to successfully disrupt 911 services, if not mitigated. However, considering that $A^*$ type bots cannot be mitigated with conventional countermeasures (e.g., blacklisting), $A^*$ bots pose a far more significant threat that should be addressed.

| | | | ***Deployed in*** | | | | ***Relevancy*** | | |
|---|---|---|---|---|---|---|---|---|---|
| | | Change in current FCC Regulations | Mobile Device | Cellular Network | E911 Network | PSAP | $NA$ | $A$ | $A^*$ |
| *Preventa* | *Disallow NSI calls* | X | | X | | | | X | X |
| | *Trusted Device Identification* | | X | X | | | | | X |
| | *Human Presence Detection* | X | X | X | | | X | X | X |
| *Mitigative* | *Blacklist DDoS Callers* | X | | X | X | | X | X | |
| | *Call Firewall* | | X | | | | X | X | X |
| | *Priority Queuing* | | | | X | X | | X | X |
| | *Silence Detection* | | | X | | | X | X | X |
| | *Disable Cell/Sector Service* | X | | X | | | X | X | X |
| | *Locate & Collect DDoS Devices* | | | | | X | X | X | X |

**Table 6. Summary of countermeasures, where those in bold are based on existing methods.**

## 7. COUNTERMEASURES

In this section we briefly survey the known countermeasures which affect $NA$ and $A$ type bots. In particular, we propose countermeasures that are specific to mitigating the threat of $A^*$ type bots. In Table 6 we summarize the findings. In general, countermeasures which are deployed earlier in the call connection process (i.e., the left of Fig. 1 in section 3) are preferable since they minimize the load on PSAPs and consume fewer resources in the network.

**Attack Prevention.** In order to prevent an attack from anonymous type bots, we propose two methods: *Disallow NSI calls* and *Trusted Device Identification*. In *Disallow NSI calls*, 911 calls from NSI devices are disallowed and no longer forwarded by wireless carriers. This method requires a change in current FCC regulations [57], and poses an ethical problem since there are those who rely of NSI devices [55]. In *Trusted Device Identification*, the device is forced to send a trusted unaltered identifier to the network. The identifier such as IMEI, must be stored in a trusted memory region (e.g., ARM TrustZone [56]) so it cannot be changed by malware at any level. Trusted device identification is used in mobile-payment solutions.

An existing preventative measure that effects anonymous type bots is *Human Presence Detection*. This method has been discussed in [57] [62]. The general approach is to have the network or PSAP automatically detect human activity associated with the 911 call. For example, the pressing numbers, DTMF pressing patterns [58], or some other captcha [59]. This method may still lead to an overload in the network if there are too many bots.

**Attack Mitigation.** If the preventative measures cannot be taken, then mitigative measures can be deployed to minimize the impact of the attack. One approach is to implement on each device a mandatory *Call Firewall*. In this approach trusted low-level software components are used to identify and block DDoS activities (e.g. frequent 911 calls), similar to what has been proposed in [60] for PCs. Another option is that each PSAP implement *Priority Queues* where callers with more reliable identifiers (e.g., valid IMSI versus some IMEI) receive higher priorities when being connected to a call-taker. This approach is only effective if the queue length is sufficiently long, however PSAPs typically only have a few extra trunks, if any at all.

To evaluate *Call Firewall* and *Priority Queues* against an $A^*$ we implemented both countermeasures in the DES. We found that *Call Firewall* was the most effective since it minimizes the load on the network and the consumption of PSAP trunks. However, this solution must be implemented in a trusted layer of the mobile phone.

There are several known and deployed countermeasures. For example, one approach is to block callers who abuse 911 (e.g. prank callers) by implementing and enforcing a *Blacklist DDoS of Callers* [62]. In this approach callers are blocked at either the network or PSAP by marking their identifiers in a shared database similar [61]. Enforcment at the entry to the network minimizes the risk of an overload. This method poses an ethical issue since will prevent the device's legitimate owner from making a 911 call [62]. Moreover, it is ineffective against $A^*$ type bots since they randomize their identifiers. Another approach is *Silence Detection* where 911 calls with no audio are blocked in the network or at the PSAPs. This approach is weak to bots that can inject recorded/synthesized audio into the call. Moreover, it poses an ethical issue since it incurs delays and is not friendly to the deaf community's needs [62].

As a last resort, law enforcement can *Locate & Collect the DDoS Devices*. This approach is not effective because locating a device is a joint effort between the police and the PSAP staff that can take anywhere between 30 minutes and 30 hours [63] requiring a lot of the police and PSAP staff's time. We estimate that it would take NC more than a week to capture the majority of an attack based on 6,000 bots. To mitigate the load of the bot traffic on the network, operators can selectively *Disable Cell/Sector Service*. Of course, this approach is undesirable, but it may benefit the greater good until the infected devices have been located or effectively blocked [62].

## 8. CONCLUSION

The threat of a DDoS attack on 911 services launched from a mobile phone botnet has not been investigated in the past. In this paper we expose a types of DDoS attack on 911 that cannot be blocked though conventional means. We show that a bot placed within the baseband firmware of a mobile phone can alter the internal protocol stack and render the device to have no genuine identification within the 2G, 3G, and 4G cellular networks. Such a bot can issue repeated emergency calls that cannot be blocked, technically or legally, by the network or the emergency call centers. We demonstrated the feasibility of this type of attack by successfully implementing and testing the bots within a small cellular network. By simulating attacks on current 911 infrastructure, we found that just 6K bots are sufficient to significantly compromise the availability of a state's 911 services (and the deployment of only 200K bots can jeopardize services across the entire US). Lastly, we enumerated device-level and network-level countermeasures and examined their effectiveness. We believe that the contributions of this paper will assist the respective organizations, lawmakers, and security professionals in understanding the scope of this issue and aid in the prevention of possible future attacks on the 911 emergency services.

APPENDIX A. NC's PSAPs SR map

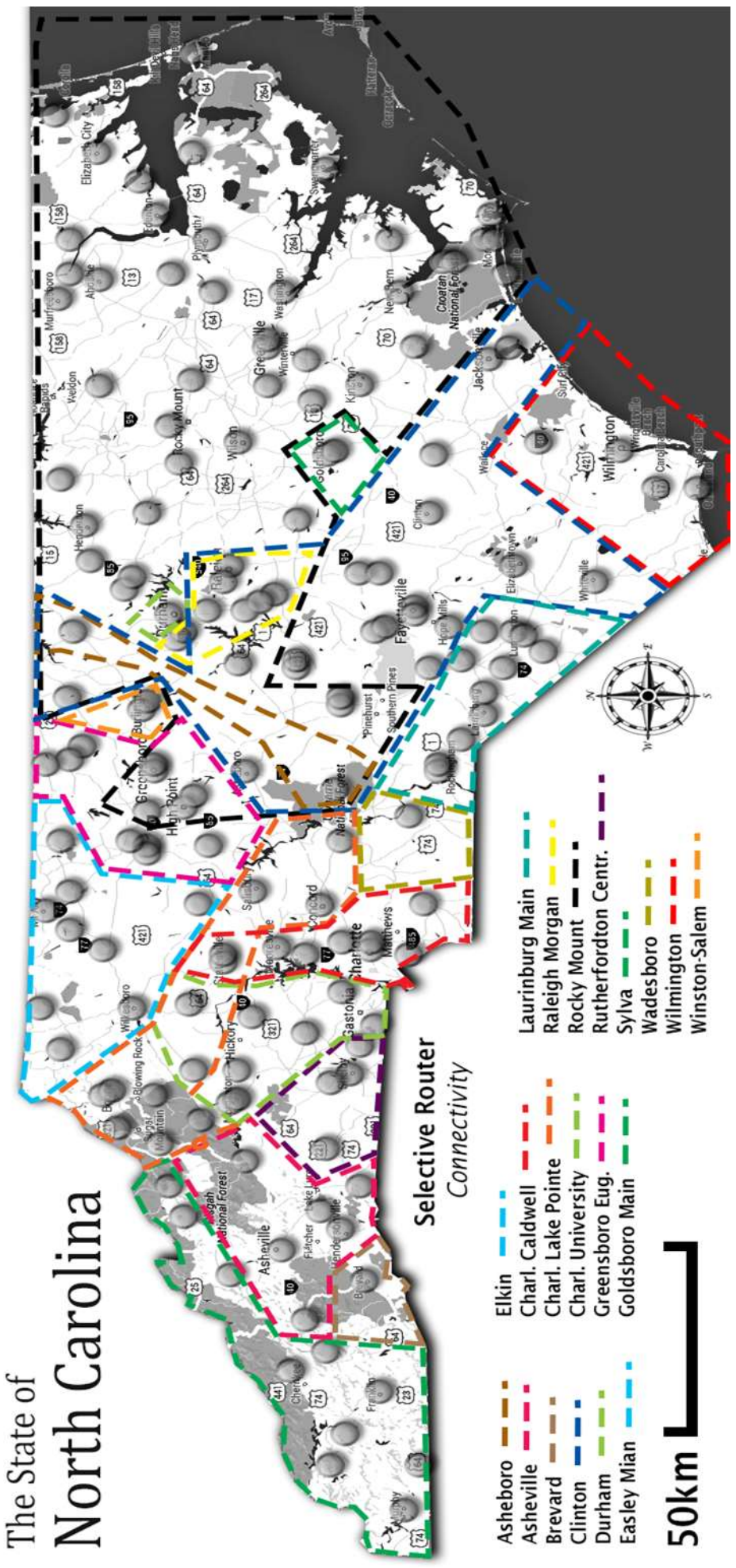

**NC's PSAPs on a map that shows which PSAPs are connected to which SR by circling them accordingly. If a PSAP is in an overlapping region, then it is an indication that the PSAP has shared responsibilities between the two geographical areas.**